\documentclass[pdflatex,sn-vancouver-num]{sn-jnl}% Vancouver Numbered Reference Style
\usepackage{graphicx}%
\usepackage{multirow}%
\usepackage{amsmath,amssymb,amsfonts}%
\usepackage{amsthm}%
\usepackage{mathrsfs}%
\usepackage[title]{appendix}%
\usepackage{xcolor}%
\usepackage{textcomp}%
\usepackage{manyfoot}%
\usepackage{booktabs}%
\usepackage{algorithm}%
\usepackage{algorithmicx}%
\usepackage{algpseudocode}%
\usepackage{listings}%
\theoremstyle{thmstyleone}%
\theoremstyle{thmstyletwo}%

\theoremstyle{thmstylethree}%

\begin{document}

\title[Dynamic Prediction for Hospital Readmission in Patients with Chronic Heart Failure]{Dynamic Prediction for Hospital Readmission in Patients with Chronic Heart Failure}


\author[1]{\fnm{Rebecca} \sur{Farina}}
\author[2]{\fnm{Francois} \sur{Mercier}}
\author[3]{\fnm{Christian} \sur{Wohlfart}}
\author[4]{\fnm{Serge} \sur{Masson}}
\author*[2]{\fnm{Silvia} \sur{Metelli}}

\affil[1]{Department of Statistics \& Data Science, Carnegie Mellon University, Pittsburgh, USA}
\affil[2]{F. Hoffmann-La Roche AG, Basel, Switzerland}
\affil[3]{Roche Diagnostics GmbH, Penzberg, Germany}
\affil[4]{Roche Diagnostics International Ltd, Rotkreuz, Switzerland}

%\affil[1]{\orgdiv{Department}, \orgname{Organization}, \orgaddress{\street{Street}, \city{City}, \postcode{100190}, \state{State}, \country{Country}}}


\abstract{
\textbf{Background:} Hospital readmission among patients with chronic heart failure (HF) is a major clinical and economic burden. Dynamic prediction models that leverage longitudinal biomarkers may improve risk stratification over traditional static models. This study aims to develop and validate a joint model using longitudinal N-terminal pro-B-type natriuretic peptide (NT-proBNP) measurements to predict the risk of rehospitalization or death in HF patients. 

\textbf{Methods:} We analyzed real-world data from the TriNetX database, including patients with an incident HF diagnosis between 2016 and 2022. The final selected cohort included 1,804 patients. A Bayesian joint modeling framework was developed to link patient-specific NT-proBNP trajectories to the risk of a composite endpoint (HF rehospitalization or all-cause mortality) within a 180-day window following hospital discharge. The model's performance was evaluated using 5-fold cross-validation and assessed with the Integrated Brier Score and Integrated Calibration Index.

\textbf{Results:} The joint model demonstrated a strong predictive advantage over a benchmark static model, particularly when making updated predictions at later time points (180-360 days). A joint model trained on patients with more frequent NT-proBNP measurements achieved the highest accuracy. The main joint model showed excellent calibration, suggesting its risk estimates are reliable.

\textbf{Conclusion:} Our findings suggest that modeling the full trajectory of NT-proBNP with a joint modeling framework enables more accurate and dynamic risk assessment compared to static, single-timepoint methods. This approach supports the development of adaptive clinical decision-support tools for personalized HF management.}




\maketitle

\section{Background}\label{sec1}
Chronic heart failure (HF) represents a significant challenge in contemporary healthcare systems due to its high prevalence, disease burden, and frequent hospital readmissions. HF affects approximately 2\% of the adult population worldwide, with prevalence rising to nearly 10\% among individuals over 75 years old \cite{bui2011epidemiology, metra2017heart}. The condition is characterized by impaired cardiac function leading to symptoms such as dyspnea, fatigue, and fluid retention \cite{ponikowski20162016}. Despite advances in therapy, HF remains associated with substantial morbidity and mortality. Notably, up to 50\% of patients experience hospital readmission within 180 days following discharge \cite{desai2012rehospitalization}, and the early post-discharge period represents a clinically vulnerable phase marked by disproportionately high event rates~\cite{greene2015vulnerable}. Accurate prediction of hospital readmission is crucial for improving clinical outcomes and reducing healthcare costs, highlighting the need for advanced predictive methodologies.

Dynamic prediction through joint modeling (JM) has gained considerable attention as a powerful statistical framework that simultaneously analyzes longitudinal biomarkers and time-to-event data. A shared random effects joint model links a survival process with a longitudinal process via common latent variables, typically random effects, to capture individual-specific trajectories and their influence on event risk. Early foundational work on joint models focused on a single longitudinal outcome modeled via linear mixed-effects models \cite{Wulfsohn1997, hogan1997}. Building on this work, subsequent research  expanded the methodology to incorporate Bayesian estimation techniques, which allow flexible inference and dynamic prediction \cite{rizopoulos2011dynamic, rizopoulos2012joint, barrett2017dynamic, andrinopoulou2018improved, ferrer2019individual, andrinopoulou2021reflection}. These frameworks have been further extended to handle more complex data structures, including competing risks \cite{andrinopoulou2014,andrinopoulou2017}, and multiple longitudinal biomarkers \cite{rizopoulos2011bayesian, mauff2020joint}. The availability of open-source software has been pivotal in applying joint models in practice. Notably, the \texttt{JM} package \cite{rizopoulos2010jm} in \texttt{R} was among the first tools for fitting joint models, while the more recent \texttt{JMbayes2} package \cite{rizopoulos2025jmbayes2} offers Bayesian joint models with enhanced flexibility.

In the context of heart failure, dynamic prediction models utilizing JM frameworks have shown great promise by leveraging longitudinal biomarker data to update risk assessments over time. Central to these models is N-terminal pro-B-type natriuretic peptide (NT-proBNP), a cornerstone biomarker released from cardiac myocytes in response to increased wall stress \cite{maisel2002rapid, felker2006natriuretic}. Higher levels of NT-proBNP reflect ongoing myocardial strain and are strongly correlated with disease severity and mortality risk \cite{maisel2002rapid, felker2006natriuretic}. 
While NT-proBNP is widely used clinically, most prognostic models rely on a single baseline measurement, which may overlook the evolving risk as disease progresses and treatments change. Indeed, the prognostic importance of changes in NT-proBNP over time, not just its baseline value, was demonstrated in major clinical trials and subsequent meta-analyses \cite{masson2008prognostic, zile2016prognostic}. This finding provides a strong rationale for more sophisticated analytical methods. A few studies have applied joint models to HF datasets, integrating longitudinal NT-proBNP measurements to generate dynamically updated risk predictions for adverse clinical events \cite{brankovic2018, vanBoven2018, zhang2018, fuery2024prognostic}. Joint models offer a powerful approach to capture these dynamic risk patterns and therefore improve individualized prognosis. Beyond HF, joint models have also been effectively applied to other cardiovascular conditions. For example, JM-based dynamic prediction has been extended to congenital heart disease data, leveraging NT-proBNP to improve prognostic accuracy \cite{baggen2018}. More recently, the field has advanced through the incorporation of competing risks into JM dynamic predictions for HF, further enhancing model applicability \cite{petersen2025}. These advances underscore the potential of dynamic prediction models to offer individualized, time-adaptive risk predictions that can support tailored clinical decision-making.

Our study builds on the existing literature by developing a robust JM framework to predict hospital readmissions in patients with chronic heart failure. Specifically, we investigate both individual and population-level longitudinal dynamics of NT-proBNP in relation to HF-related rehospitalization, evaluate whether patient-specific NT-proBNP trajectories from hospital discharge to readmission can dynamically predict clinical outcomes, and identify the most informative functional form linking biomarker evolution to event risk. Indeed, whether absolute levels or relative changes in NT-proBNP serve as more important drivers of risk is an aspect that remains unresolved in previous studies \cite{brankovic2018,zhang2018}. By leveraging dynamic prediction, we intend to better characterize patient-specific evolution and empower clinicians with adaptive tools for optimizing patient management.

\section{Methods}\label{sec2}
\subsection{Data preprocessing and filtering}
This study utilized real-world data from TriNetX heart (TriNetX, LLC; \href{https://live.trinetx.com}{https://live.trinetx.com}), comprising 400,003 patients with incident heart failure diagnoses between 2016 and 2022. The data covers 49 healthcare organizations within the United States and captures demographics information, other diagnoses, laboratory results and vital signs. To ensure robustness and minimize biases due to missing information, the analysis was restricted to patients with complete records. All patients underwent routine follow-up care at outpatient clinics by treating physicians who were blinded to biomarker sampling and results. Clinical events such as HF hospitalization, myocardial infarction, and mortality were recorded in electronic case report forms, supported by hospital records and discharge letters. 
%A blinded clinical event committee adjudicated these endpoints by thorough review of the collected documents.
We employed specific ICD-10-CM diagnosis codes and discharge status codes to ascertain these endpoints.
During our analysis, important inconsistencies were observed in death dates and admission-discharge visit dates. For example, some death dates preceded or implausibly overlapped with hospitalization periods, and some admission dates occurred after the corresponding discharge date. These discrepancies were addressed by discarding such patients in order to maintain high data quality.

\subsubsection{Endpoint definition and estimand of interest}
For each patient, the landmark time $t_0$ (or index time) is defined as the date of the hospital discharge due to the first heart failure diagnosis. The event of interest is a composite endpoint consisting of either rehospitalization due to HF or death from any cause, whichever occurs first. The time-to-event variable measures the duration from $t_0$ to the occurrence of the event, if present. If the event does not occur during the follow-up period, the patient is considered right-censored. In this case, the censoring time measures the duration from $t_0$ to the discharge date of the patient's last follow-up visit. A key limitation in the data is that around 52\% of censored patients have no documented follow-up visits after their HF hospitalization. As a result, they were excluded from the analysis. Roughly  25\% of the total initial cohort was retained, having no date inconsistencies and with a documented event or follow-up visit. The primary quantity of interest is the individual-specific risk of experiencing the composite event within a prediction window of 180 days following the landmark time $t_0$. This dynamic risk prediction aims to support timely clinical decision-making by providing updated prognostic information tailored to each patient’s evolving clinical trajectory.

\subsubsection{Longitudinal Biomarker: NT-proBNP}
The primary longitudinal biomarker analyzed in this study is NT-proBNP, a well-established indicator strongly associated with heart failure outcomes \cite{brankovic2018, vanBoven2018, zhang2018, petersen2025}. To be included in the analysis, patients were required to have at least one NT-proBNP measurement within 30 days before their first HF hospitalization ($t_0$) and a minimum of two measurements after $t_0$ but before the occurrence of the event or the last follow-up visit. These conditions ensure sufficient longitudinal data to capture individual biomarker trajectories. In cases where multiple NT-proBNP measurements were recorded on the same date for a patient, the average value is used to represent that day. Furthermore, values of NT-proBNP exceeding 70,000 pg/mL were excluded from the analysis based on clinical expertise, as such extreme levels may indicate measurement anomalies. After applying these NT-proBNP related inclusion criteria (following prior exclusions for data inconsistencies and follow-up loss), approximately 8\% of the initial cohort remained eligible for our study.

\subsubsection{Baseline characteristics}
Our analysis includes baseline characteristics such as demographic information, vital signs, comorbidities, and laboratory test results. Although the TriNetX dataset contains a wide range of features, we focus on the most relevant to HF, as determined by prior research and clinical expertise. Moreover, to ensure robustness, only complete cases (patients without missing values in all selected features) are included. Thus, prior to statistical analysis, we identified a set of baseline covariates that balances clinical relevance and data completeness, thereby enhancing model reliability. The selected baseline characteristics include gender, age, body mass index (BMI), creatinine, systolic blood pressure (SBP), atrial fibrillation (AFib), anemia, chronic kidney disease (CKD), chronic obstructive pulmonary disease (COPD), diabetes mellitus, hypertension, myocardial infarction (MI), and stroke. Notably, some clinically important variables could not be included due to excessive missing values or absence from the TriNetX database: ethnicity, New York Heart Association (NYHA) functional class, heart rate (HR), and left ventricular ejection fraction (LVEF). The features are incorporated exclusively in the survival component of the joint model, and are treated as time-invariant, measured at baseline. More precisely, we do not require the covariates to be recorded at the landmark time $t_0$, but they must have been collected within 30 days prior to $t_0$ to ensure clinical relevance.
%After applying the complete-case selection criteria, roughly 0.5\% of the initial cohort remains eligible (400,003 patients), resulting in a final sample of 1,804 patients. 
We emphasize that the selection of baseline characteristics presented here was determined a priori, based on medical relevance. While feature selection is an important aspect of predictive modeling, it is beyond the scope of the current study and is reserved for future research. Data-driven variable selection techniques, such as LASSO, elastic net regularization, random survival forests, or gradient boosting, could be applied in preliminary survival analyses to identify the most predictive covariates. After applying the complete-case selection criteria, the final analysis cohort consisted of 1,804 patients. The corresponding baseline characteristics and their summary statistics are reported in Table \ref{tab:covariates}. 
Continuous covariates were standardized prior to statistical analysis to facilitate model convergence. 
\begin{table}[t]
\caption{Summary of baseline demographic and clinical characteristics for the study cohort (n=1,804).}
\label{tab:covariates}
\centering
\begin{tabular}{lr}
    \toprule 
    \textbf{Characteristic} & mean (SD) / n (\%) \\ 
    \midrule 
    Age (years) & 67.1 (12.9) \\
    Gender (male) & 952 (53) \\
    Body Mass Index (kg/m$^2$) & 32.4 (9.0) \\ 
    Systolic Blood Pressure (mmHg) & 131.6 (26.1) \\
    Creatinine (mg/dL) & 1.4 (1.0) \\
    Atrial fibrillation & 700 (39) \\
    Anemia & 606 (34) \\
    Chronic kidney disease & 439 (24) \\
    Chronic obstructive pulmonary disease & 546 (30) \\
    Diabetes & 765 (42) \\
    Hypertension & 1,358 (75) \\
    Myocardial infarction & 432 (24) \\
    Stroke & 292 (16) \\
    \bottomrule
\end{tabular}
\footnotetext{Continuous variables are presented as mean (standard deviation), while categorical variables are shown as counts (percentages). All features were measured within 30 days prior to the landmark time (date of first HF hospitalization discharge).}
\end{table}

All selection criteria were essential to ensure high data quality and consistent longitudinal information, which is a prerequisite for reliable joint modeling of biomarker trajectories. As a consequence, the final cohort should not be viewed as representative of the original population in the TriNetX dataset. The goal of this study, however, is not to derive population-level or generalizable epidemiologic conclusions, but rather to provide a methodological proof of concept: specifically, to demonstrate that leveraging full longitudinal NT-proBNP trajectories within a joint modeling framework can substantially enhance dynamic prediction of HF outcomes.

\subsection{Statistical methods}
\subsubsection{Joint modeling of longitudinal and survival data}
The JM framework offers a powerful approach to simultaneously analyze the evolution of a longitudinal biomarker and its direct impact on the time to a clinical event, while properly accounting for measurement error and individual-level variability. Let $T_i$ denote the observed event time for the $i$-th individual, with $i = 1, \dots, n$, defined as the minimum between the true event time $T_i^*$ and the censoring time $C_i$, i.e., $T_i = \min(T_i^*, C_i)$. The event indicator is defined as $\delta_i = I(T_i^* \leq C_i)$. The longitudinal biomarker measurement (NT-proBNP) for individual $i$ at time $t$ (in days) is denoted by $y_i(t)$. The observed longitudinal data for individual $i$ consists of measurements $y_i = \{y_i(t_{ij})\}_{j=1}^{n_i}$ collected at times $t_i = \{t_{ij}\}_{j=1}^{n_i}$, where $n_i$ is the number of observations for that individual. Hence, the complete observed data for subject $i$ is $(T_i, \delta_i, y_i, t_i)$. We denote by $m_i(t)$ the true and unobserved value of the longitudinal outcome at time $t$. To assess its effect on event risk, we estimate $m_i(t)$ via a mixed-effects model, which combines fixed effects to capture population-level trends and random effects to characterize individual variation, a well-established approach in longitudinal data analysis \cite{laird1982random, verbeke2000linear}. Specifically, we incorporate quadratic splines of time to flexibly model subject-specific trajectories:
\[
y_i(t) = m_i(t) + \epsilon_i(t) = \left( \beta_0 + \sum_{v=1}^{2} \beta_v B_v(t) \right) + \left( b_{0i} + \sum_{v=1}^{2} b_{vi} B_v(t) \right) + \epsilon_i(t),
\]
%linear
%\begin{align*}
%   m_i(t) &= \left( \beta_0 +  \beta_1t \right) + \left( b_{0i} + b_{1i} t \right) 
%\end{align*}
where $B_v(t)$ are the basis functions for quadratic splines, $\beta_0, \beta_1, \beta_2$ are fixed-effects coefficients representing the population-average trajectory, $b_{0i}, b_{1i}, b_{2i}$ are the subject-specific random effects, and $\epsilon_i(t)$ is the measurement error at time $t$. The random effects vector $b_i = (b_{0i}, b_{1i}, b_{2i})^\top$ is assumed to be independent of the error terms and to follow a multivariate normal distribution:
\[
b_i \sim \mathcal{N}(0, \Sigma_b).
\]
Moreover, the measurement errors are assumed to be independent between subjects and normally distributed:
\[
\epsilon_i(t) \sim \mathcal{N}(0, \sigma^2).
\]

In joint modeling, to link the longitudinal biomarker with the risk of the event, a hazard model is specified. The hazard rate for individual $i$ at time $t$, denoted $h_i(t)$, is assumed to depend on the true underlying longitudinal value $m_i(t)$ and the vector of baseline covariates $x_i$:
%A commonly used form for this hazard model assumes that the instantaneous risk of an event at time $t$ is directly influenced by the current value of the true longitudinal process:
\[
h_i(t) = h_0(t) \exp\left\{ \gamma^\top x_i + \alpha m_i(t) \right\},
\]
where $h_0(t)$ is the baseline hazard function, $\gamma$ represents the coefficients for the baseline covariates, and $\alpha$ quantifies the association between the biomarker's current value and the hazard of the event.

%$g(\cdot)$ is a link function that transforms the true longitudinal value $m_i(t)$ to connect it to the hazard scale, 

Crucially, the hazard depends on the true unobserved longitudinal process $m_i(t)$ rather than the noisy observations $y_i(t)$. This connection is established through the random effects $b_i$, which serve as the bridge between the longitudinal and survival submodels, capturing how individual-specific biomarker trajectories relate to their risk of experiencing the event. For the baseline hazard function, several specifications are possible. In this study, $h_0(t)$ is modeled parametrically using the Weibull distribution \cite{kalbfleisch2002statistical}. Alternatively, one could use other distributions or leave $h_0(t)$ unspecified, as in the Cox proportional hazards regression.

%The link function $g(\cdot)$ defines how the longitudinal marker influences the hazard. Common choices for include the estimated true value of the biomarker at time $t$ for individual $i$, the estimated true value of the biomarker at baseline ($t=0$) for individual $i$, the estimated true value of the biomarker at the time of the event for individual $i$, or the estimated individual-specific slope of the biomarker (representing a time-invariant effect of the longitudinal trend).

A classical approach to estimating joint model parameters relies on maximum likelihood, where the log-likelihood is derived from the joint distribution of the longitudinal and time-to-event outcomes. 
%Under the assumption that subject-specific random effects $b_i$ drive both processes, the conditional likelihood for subject $i$ factorizes as:
%\[
%p(T_i, \delta_i, y_i \mid b_i; \theta) = p(T_i, \delta_i \mid b_i; \theta) \cdot \prod_j p(y_i(t_{ij}) \mid b_i; \theta),
%\]
%where $\theta$ is the full parameter vector. 
Obtaining the marginal likelihood requires approximation due to intractable integrals. Methods like Gaussian quadrature or Monte Carlo integration are commonly used \cite{henderson2000joint, song2002semiparametric}. In this work, we instead adopt a Bayesian approach \cite{rizopoulos2011dynamic, rizopoulos2012joint}, where inference is based on the posterior distribution of the model parameters and latent variables. Specifically, we use Markov Chain Monte Carlo (MCMC) methods to estimate the parameters of the proposed joint model. The joint likelihood is formulated under the assumption that the longitudinal and survival processes are conditionally independent given the subject-specific random effects.

For posterior sampling, the random effects $b_i$ are updated using a Metropolis-Hastings algorithm. Uncertainty quantification is facilitated by summarizing the posterior samples via standard errors or credible intervals derived from quantiles of the Monte Carlo draws. Overall, the Bayesian MCMC approach allows for flexible incorporation of prior knowledge and provides a natural framework for uncertainty quantification in joint models.

\subsubsection{Model specification}
The longitudinal submodel, describing the evolution of log-transformed NT-proBNP values over time, was implemented via the \texttt{nlme} \cite{pinheiro2017package} package. To flexibly capture the non-linear subject-specific trajectories and variability over time (see Figure \ref{fig:EDA}F), we used quadratic splines of time for both fixed and random effects. We did not include additional covariates in the longitudinal model, in order to focus on the temporal dynamics of NT-proBNP and its association with event risk, and to avoid overfitting given the limited number of measurements for many patients. The survival submodel includes all pre-selected baseline covariates. We assumed a parametric Weibull baseline hazard to enable fully parametric estimation of the time-to-event distribution, implemented using the \texttt{survival} package \cite{therneau2015package}. 

We implemented joint models to capture the relationship between the NT-proBNP trajectories and the risk of heart failure-related rehospitalization or death. 
The survival risk is linked to the longitudinal process through the estimated current value $m_i(t)$ of the biomarker. We explored different functional forms linking the longitudinal trajectory with the survival risk, including the slope of the biomarker (i.e., instantaneous rate of change), and the combination of the current value and the slope. Preliminary results indicated that the current biomarker value alone provided the best trade-off between predictive performance and model stability, likely due to the limited number and irregular timing of biomarker measurements. 

Modeling a dynamic longitudinal process is particularly challenging when data are sparse at patient level; to assess robustness and examine the impact of the sparsity of NT-proBNP measurements on model performance, we fitted two variants of the joint model:
\begin{itemize}
\item A main model, using the 1,804 patients selected from the initial cohort according to our inclusion criteria. 
%\item A restricted-time model, in which NT-proBNP measurements and event times were truncated at 2 years (720 days) after $t_0$, to focus on more immediate clinical trajectories. We filter patients who have at least two biomarker values in this time window (and at least one value prior to or at $t_0$), resulting in a smaller sample size of 1307.
\item A high-frequency sampling model, using only patients with at least five post-$t_0$ NT-proBNP measurements (and at least one value before or at $t_0$). This subgroup allowed for more accurate trajectory estimation but was too small (369 patients) to be considered as the primary analysis set.
\end{itemize}
For benchmarking purposes, we also fitted a standard survival Weibull model using all baseline covariates and the most recent NT-proBNP measurement as an additional covariate, consistent with the approach in prior work~\cite{zhang2018}, and refer to this model as the benchmark model. This comparison is intended to determine whether modeling the entire biomarker trajectory dynamically offers advantages over relying on a single static measurement.
All models were estimated using the \texttt{JMbayes2}\cite{rizopoulos2025jmbayes2} package in \texttt{R}, which employs a full Bayesian framework with MCMC algorithms. Posterior inference is obtained via a combination of Gibbs sampling and Metropolis-Hastings algorithm. We used the default priors in the package \texttt{JMbayes2}, which primarily consist of weakly informative normal priors for the regression coefficients and half-t or gamma priors for the variance and standard deviation components.

\section{Results}\label{sec3}
\subsection{Exploratory data analysis}
\begin{figure}[t!]
    \centering
    \includegraphics[width=\textwidth]{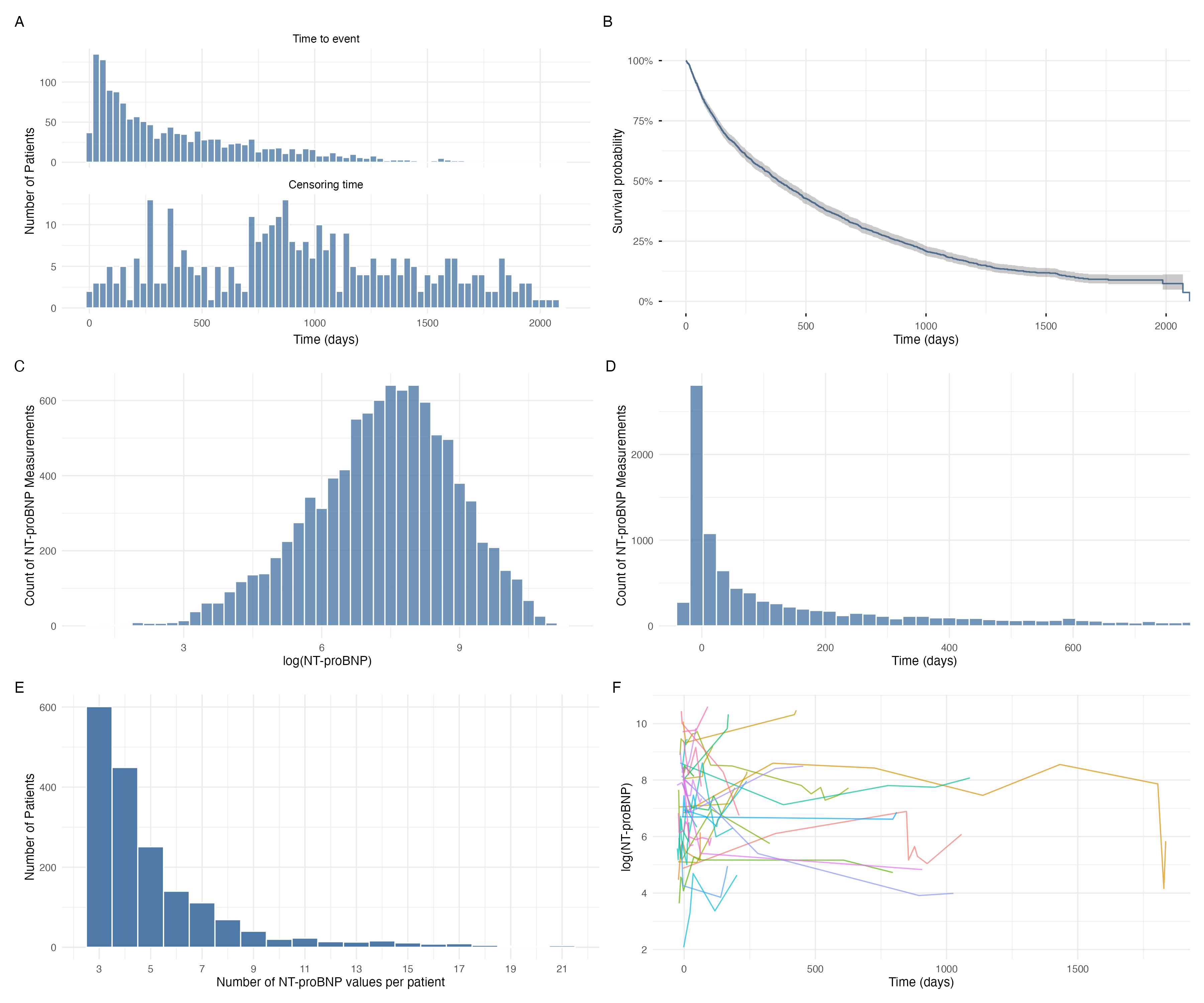}
    \caption{Exploratory data analysis of the study cohort. (\textbf{A}) Distribution of the time-to-event (top) for patients who experienced the composite endpoint (n=1,457) and distribution of censoring time (bottom) for patients who did not experience the event (n=347). (\textbf{B}) Kaplan-Meier curve for the composite endpoint (HF rehospitalization or all-cause mortality). The solid line represents the estimated probability of remaining event-free over the follow-up period, with the shaded area indicating the 95\% confidence interval. (\textbf{C}) Histogram of log-transformed NT-proBNP measurements (n=9,594) collected from any patient in the study cohort. The transformation was used to produce a more symmetric distribution for statistical modeling. (\textbf{D}) Distribution of the collection times for all NT-proBNP measurements relative to the landmark time $t_0.$ The majority of measurements were recorded early in the follow-up period. (\textbf{E}) Distribution of the number of NT-proBNP measurements per patient. A significant proportion of patients (33\%) have three recorded values, which is the minimum required by our inclusion criteria. (\textbf{F}) Individual longitudinal trajectories of log-transformed NT-proBNP levels over time for a random subset of 35 patients. Each colored line represents a single patient, illustrating the substantial inter-patient heterogeneity in biomarker patterns over time.}
    \label{fig:EDA}
\end{figure}

The study cohort, selected from the TriNetX database following the criteria described previously, includes 1,804 patients. Among them, 1,457 patients (approximately 80\%) experienced the composite event. Figure \ref{fig:EDA}A (top panel) illustrates the distribution of the time-to-event for these patients, which is highly skewed, with a median time-to-event of 267 days (IQR: 97 - 585 days). The mean time-to-event is 388 days, and although 40\% of events occur within the first 180 days, the range extends to nearly 6 years, far beyond the 180-day window of interest. The bottom panel of Figure \ref{fig:EDA}A presents the distribution of the censoring time for patients who did not experience the event, which appears more uniformly spread over the follow-up period. Additionally, the Kaplan–Meier survival curve in Figure \ref{fig:EDA}B provides a visual summary of the decreasing survival probability over time, highlighting the progressive risk of experiencing HF rehospitalization or death during the follow-up window.

The NT-proBNP measurements in this cohort span a wide range, with a median concentration of 1757 pg/mL (IQR: 578 - 4637.5 pg/mL).
In our analysis NT-proBNP values are log-transformed to ensure positivity and to better approximate a symmetric distribution, following common practice in the literature \cite{vanBoven2018, zhang2018}. Figure \ref{fig:EDA}C displays the distribution of the log-transformed NT-proBNP measurements collected from any patient, totaling 9,594 values. The time distribution of these values is shown in Figure \ref{fig:EDA}D, which illustrates that the time measurements of NT-proBNP are concentrated in the first part of the follow-up period, with almost 70\% of the values recorded within the first 180 days after $t_0$. Recall that biomarker measurements taken within 30 days prior to $t_0$ and at $t_0$ are included in the analysis; in fact, approximately 31\% of all measurements fall within this pre-baseline window. The average time of the first NT-proBNP measurement per patient is 9 days before $t_0$, while the average time of the last measurement is 367 days after $t_0$. The distribution of the number of NT-proBNP measurements per patient (Figure \ref{fig:EDA}E) indicates that the largest proportion of patients (33\%) have exactly three recorded values, which is the minimum required by our inclusion criteria. The limited number of longitudinal measurements represents a key challenge of the analysis, as it constrains our ability to robustly characterize individual biomarker trajectories over time. Individual longitudinal trajectories, shown in Figure \ref{fig:EDA}F further demonstrate substantial heterogeneity in NT-proBNP patterns.

\subsection{Model fit and baseline risk associations}
\begin{figure}[t]
    \centering
    \includegraphics[width=\textwidth]{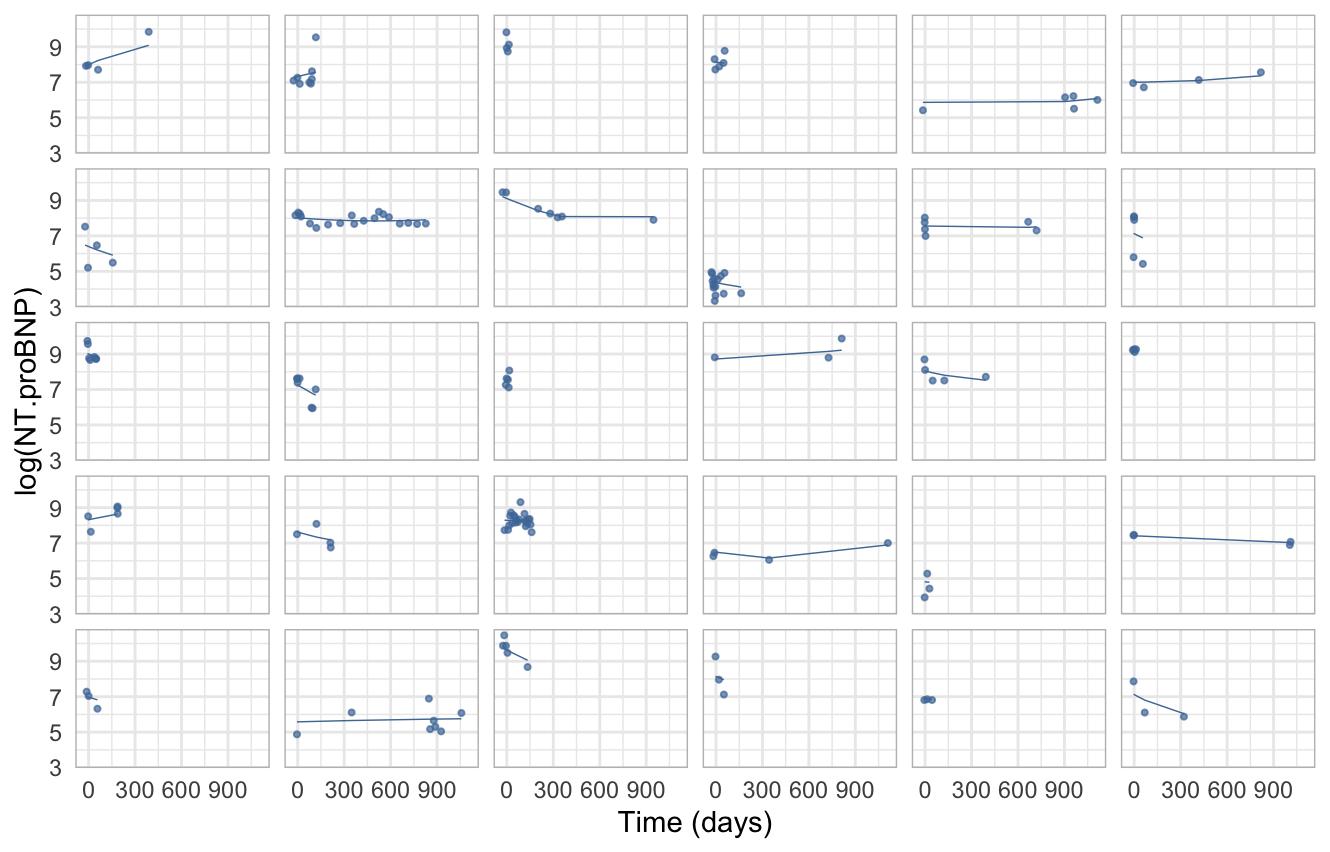}
    \caption{Fitted individual NT-proBNP trajectories from the longitudinal submodel for a random subset of 30 patients. Points are the observed log(NT-proBNP) values, and the solid lines are the corresponding fitted trajectories from the model.} 
    \label{fig:obs-pred-NT}
\end{figure}

To qualitatively assess the fit of the longitudinal submodel, Figure \ref{fig:obs-pred-NT} shows the observed log(NT-proBNP) measurements alongside the corresponding predicted individual trajectories for a randomly selected subset of patients. The fitted curves closely track individual biomarker changes over time, even in the presence of sparse and irregular measurement schedules. This visual agreement indicates that the spline-based mixed-effects structure captures both the overall population trend and subject‐specific deviations, supporting its suitability for downstream dynamic prediction.

\begin{figure}[t]
    \centering
    \includegraphics[width=\textwidth]{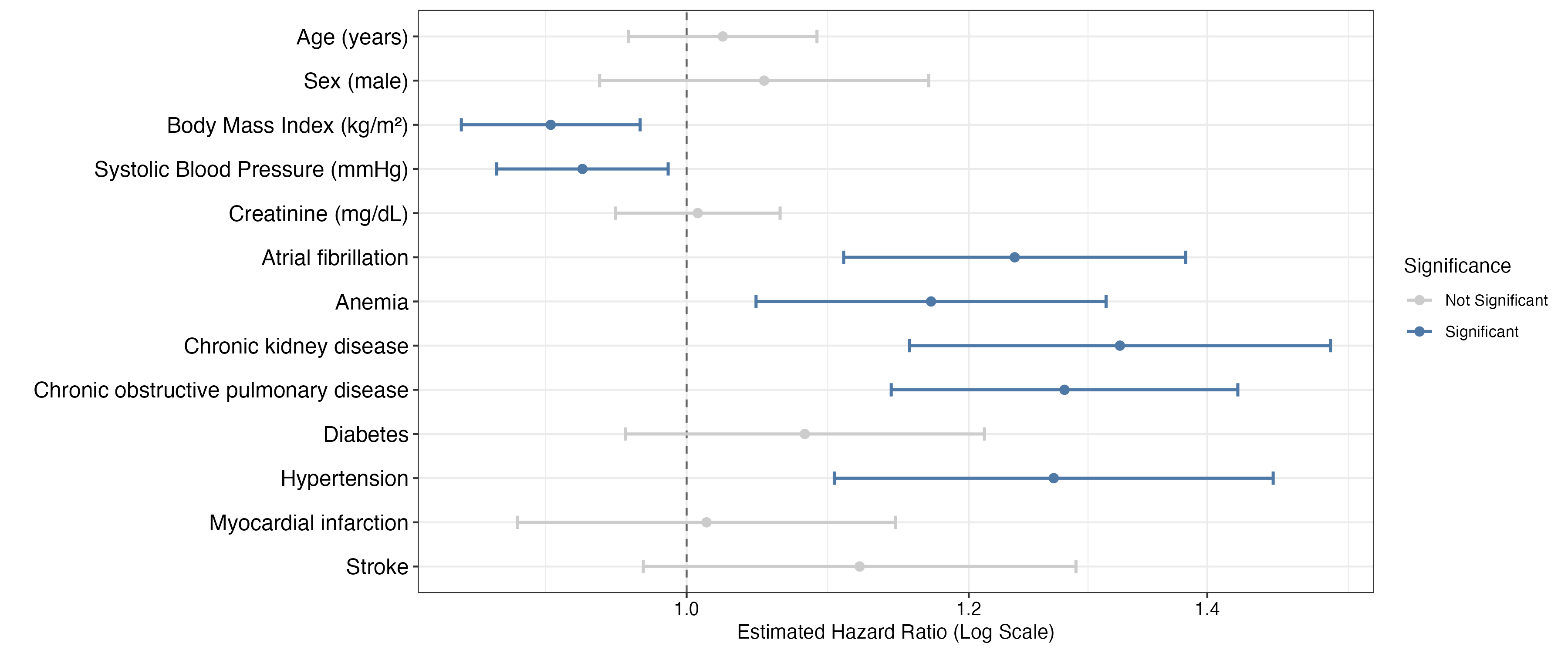}
    \caption{Forest plot of hazard ratios for baseline predictors in the survival (Weibull) submodel. The plot displays the estimated hazard ratios (dots) and their corresponding 95\% confidence intervals (horizontal bars) for the composite endpoint of heart failure rehospitalization or all-cause mortality. Each hazard ratio (HR) is derived from the Weibull model coefficients ($\beta$) and scale parameter ($\sigma$) using the transformation $\text{HR} = \exp(-\beta / \sigma)$. An HR $> 1$ indicates an increased risk of the event. Statistically significant associations, based on 95\% confidence level, are highlighted in blue.} 
    \label{fig:forest}
\end{figure}

In parallel, Figure \ref{fig:forest} summarizes the estimated associations between baseline covariates and the composite outcome using the Weibull survival model. The forest plot illustrates hazard ratios with 95\% confidence intervals, providing a benchmark static risk profile for the cohort. Each hazard ratio represents the multiplicative change in the risk of the event associated with a one-unit increase (or presence of a condition), holding all other predictors constant. Several predictors emerge as statistically significant drivers of adverse outcomes, including BMI, creatinine, systolic blood pressure, atrial fibrillation, anemia, chronic kidney disease, COPD, and hypertension. These associations align with established clinical understanding of HF prognosis~\cite{ponikowski20162016} and confirm that the baseline model captures clinically meaningful patterns before incorporating longitudinal NT-proBNP information.

\subsection{Individualized dynamic predictions}\label{Sec:indiv}
\begin{figure}[t]
    \centering
    \includegraphics[width=\textwidth]{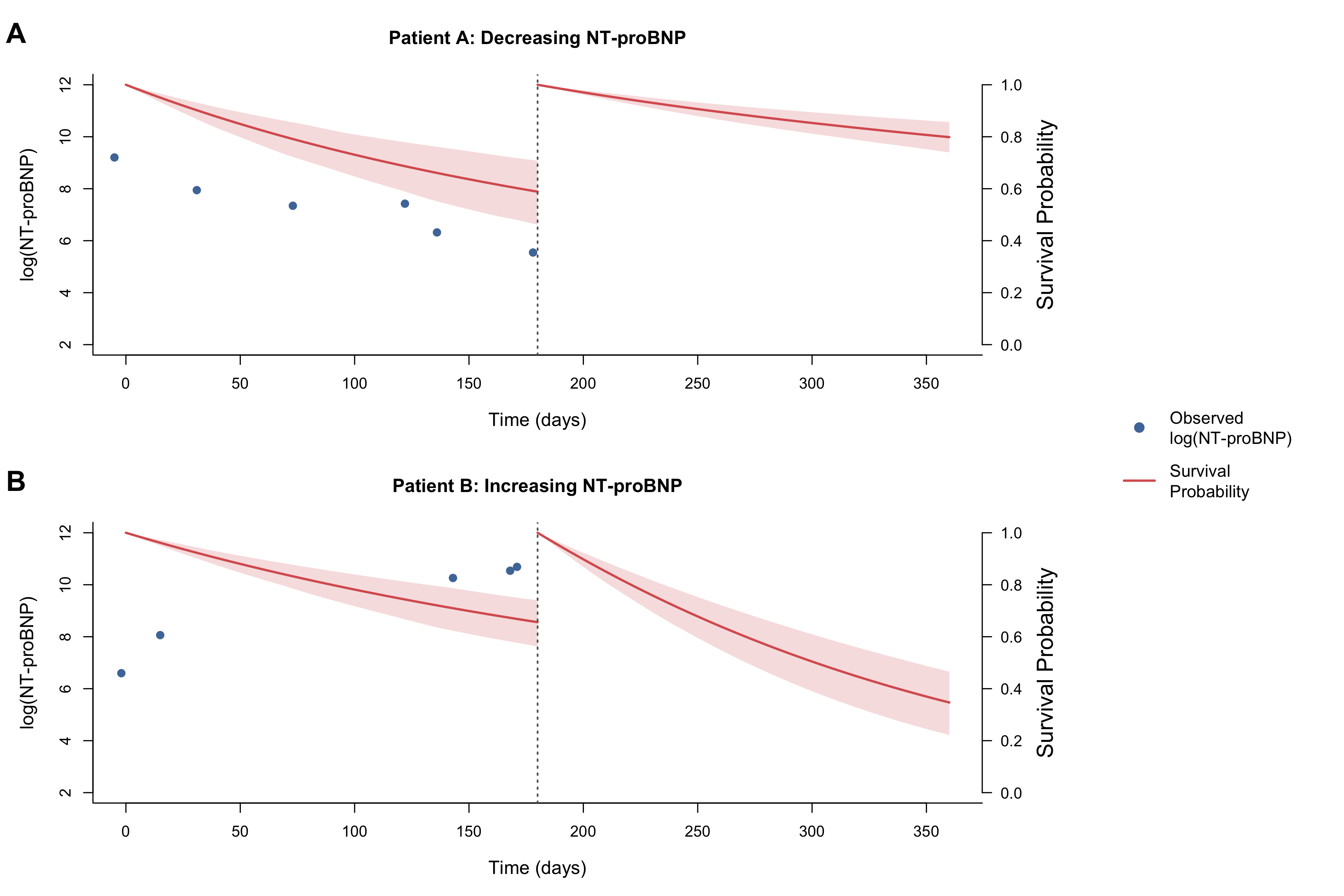}
    \caption{Individualized dynamic survival predictions for two representative patients. In both panels the left-most red curve displays the initial predicted survival probability for the 0-180 day window, while right-most red curve shows the updated prediction for the 180-360 day window, incorporating new biomarker data. The red shaded area is its relative 95\% credible interval. The blue dots represent the observed log-transformed NT-proBNP values. (\textbf{A}) Patient with a decreasing NT-proBNP trend, leading to an improved survival probability estimate at 180 days. (\textbf{B}) Patient with an increasing NT-proBNP trend, resulting in a worsened prognosis. 
    } 
    \label{fig:individ-preds}
\end{figure}

A key strength and primary advantage of the JM framework is its ability to generate dynamic, individualized predictions that are updated as new patient information becomes available. We leverage our fitted models to estimate the probability that a patient, who has so far remained event-free, will experience the composite endpoint of HF rehospitalization or death within a subsequent 180-day window. This prediction is conditioned on the patient's specific history of NT-proBNP levels and baseline characteristics, allowing for a personalized risk assessment that evolves over time. To illustrate the dynamic feature of JM, we calculate individualized survival probability predictions at two distinct time points. First, at $t_0$ we generate a prediction for a patient's probability of surviving the initial 180-day period, using the available NT-proBNP values measured before $t_0$. Then, supposing that the patient has survived these first six months, we generate a new updated prediction for the next six months (i.e., from day 180 to day 360). This second prediction leverages the additional NT-proBNP values collected during the first six months interval. This approach allows us to demonstrate how a patient's estimated risk can change in response to their evolving NT-proBNP trajectory during the initial six months of follow-up. For example, in Figure~\ref{fig:individ-preds}A, a patient whose NT-proBNP levels decrease following discharge has a lower updated risk (higher survival probability) at 180 days compared to their baseline prediction. In Figure~\ref{fig:individ-preds}B, a patient with increasing NT-proBNP levels have a lower survival probability in the next 180 days.

These findings demonstrate that the full trajectory of NT-proBNP provides critical prognostic information that is lost when using only static, single-timepoint methods. This is a key contribution of our work and stands in contrast to previous studies~\cite{zhang2018}, which concluded that the last observed NT-proBNP value may suffice for risk prediction. Our results challenge this, highlighting that risk is not static but evolves. As shown in our performance analysis, the joint model’s ability to dynamically incorporate the direction of the biomarker trend (i.e., improving or worsening) provides a more accurate and adaptive risk assessment than a static benchmark model. Our model therefore supports a more sophisticated clinical decision-making process, informed by the patient's entire biomarker trajectory rather than a single, isolated snapshot. 
%In fact, this adaptive methodology enables a more precise and individualized prognosis by capturing the patient's full clinical trajectory, a key limitation of static, single-point assessments.

\subsection{Predictive performance}\label{Sec:pred-perf}
To rigorously evaluate and validate the predictive performance of our joint models, we employed a 5-fold cross-validation procedure. The dataset was randomly partitioned into five equally-sized folds. In each of the five iterations, one fold was held out as the test set for an unbiased assessment of the model's performance, while the other four folds were combined to form the training set used to train the model. This process was repeated until every fold had been used as a test set once, ensuring that predictions for the entire dataset were generated on data not seen during model training.

The primary metric used to assess the accuracy of the risk predictions is the Integrated Brier Score (IBS). The IBS measures the mean squared difference between the predicted probabilities and the actual outcomes over time, providing a comprehensive measure of a model's overall performance that incorporates both discrimination (the ability to distinguish patients who will have an event from those who will not) and calibration (the level of agreement between predicted risks and observed event frequencies). The score ranges from 0 to 1, where a lower score indicates a more accurate model, and a score of 0.25 corresponds to a model that consistently predicts a 50\% probability for each outcome, reflecting no predictive skill. Furthermore, we calculate the Integrated Calibration Index (ICI), which specifically quantifies a model's calibration. An ICI of 0 indicates perfect calibration, where the predicted risks align perfectly with the observed outcomes at all risk levels. ICI scores ranging from approximately 0.01 to 0.02 indicate excellent calibration. An ICI value of 0.1 or higher would be regarded as poor calibration.

The aggregated performance scores from the cross-validation are presented in Tables \ref{tab:ibs} and \ref{tab:ici}, which reports the average IBS and ICI for both the training and test folds for our main joint model, the high-frequency joint model, and the benchmark standard survival model. Consistent with the predictive horizons used for the individualized dynamic predictions, these metrics are calculated for the 0-180 day and 180-360 day time windows. Reporting the results for both training and test sets allows us to assess the degree of overfitting: a large discrepancy between the two scores would indicate that a model does not generalize well to new data. The main joint model demonstrates comparable accuracy (IBS) to the benchmark within the 0–180 day horizon but outperforms it in the subsequent 180–360 day period by leveraging updated longitudinal biomarker data. This ability to dynamically incorporate new information over time is a key strength of joint models that the static benchmark model lacks. The high-frequency joint model consistently achieves the lowest IBS across both time horizons, indicating better overall prediction accuracy. The improvement in IBS from the main joint model to the high-frequency joint model is expected, as the latter benefits from more frequent NT-proBNP measurements which allow for more accurate prediction of the biomarker trajectory. Importantly, the close agreement between training and test set scores for all models suggests minimal overfitting and good generalizability. 

\begin{table}[t]
\caption{Comparison of predictive accuracy using the Integrated Brier Score (IBS) for the 0-180 and 180-360 day prediction windows.}
\centering
\begin{tabular}{l l c c}
\hline
 & & 0-180 days & 180-360 days \\
\hline
Joint Model &  Train & 0.1375 & 0.1008 \\
& Test & 0.1384 & 0.1009\\
High-Frequency JM & Train & 0.0740 & 0.0696\\
& Test & 0.0756 & 0.0730\\
Benchmark Model & Train & 0.1319 &  0.2247\\
& Test & 0.1328 & 0.2271 \\
\hline
\end{tabular}
\label{tab:ibs}
\end{table}

\begin{table}[t]
\caption{Comparison of model calibration using the Integrated Calibration Index (ICI) for the 0-180 and 180-360 day prediction windows.}
\centering
\begin{tabular}{l l c c}
\hline
 & & 0-180 days & 180-360 days \\
\hline
Joint Model &  Train & 0.0131 &  0.0221\\
& Test & 0.0177 & 0.0283\\
High-Frequency JM & Train & 0.0166 & 0.0323 \\
& Test & 0.0329 & 0.0372\\
Benchmark Model & Train & 0.0310 & 0.0283\\
& Test & 0.0422 & 0.0423\\
\hline
\end{tabular}
\label{tab:ici}
\end{table}

\begin{figure}[t]
    \centering
    \includegraphics[width=0.75\textwidth]{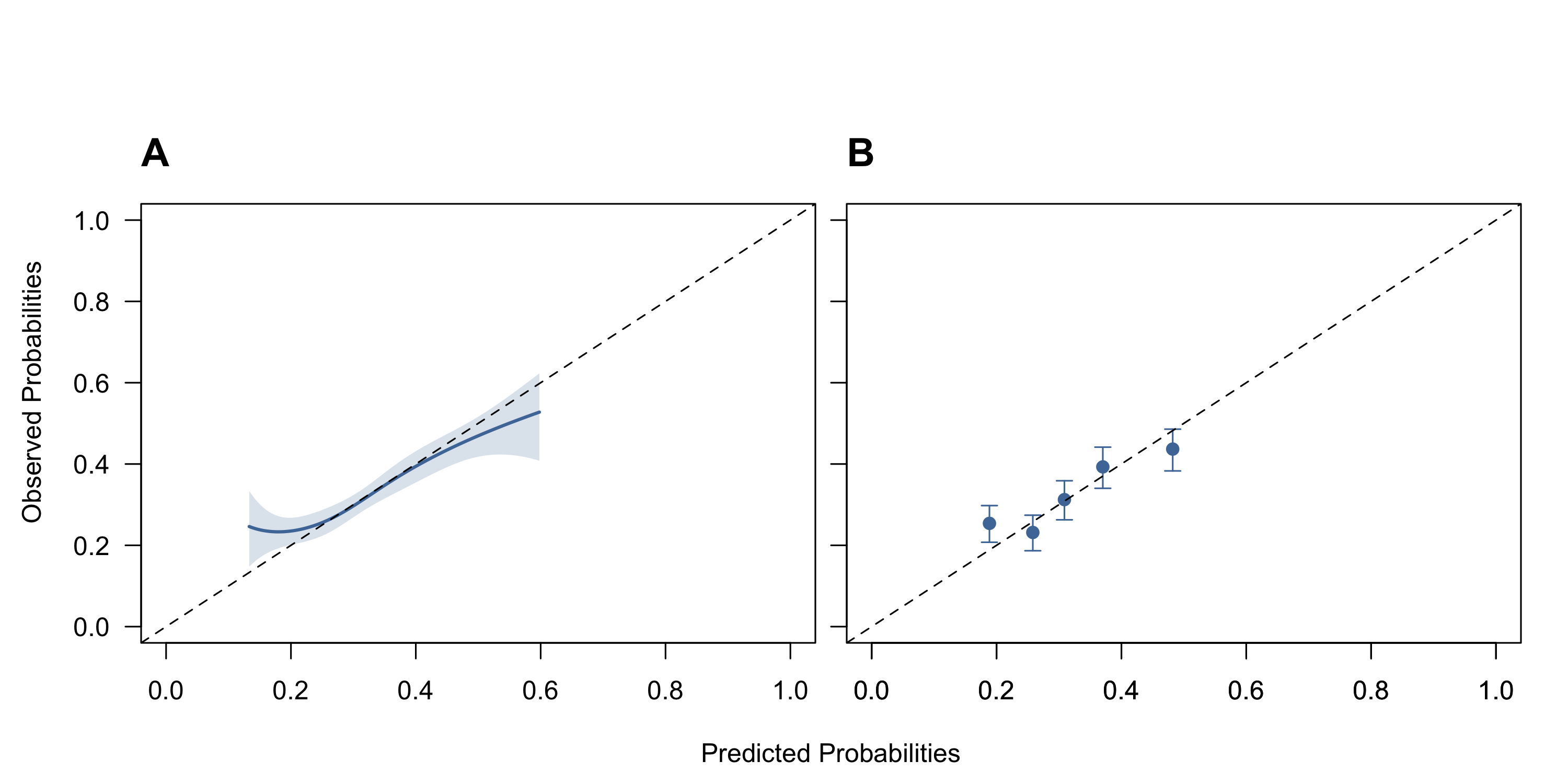}
    \caption{Calibration of the main joint model’s risk predictions for the 0–180 day period post-discharge. (\textbf{A}) Smoothed calibration curve based on a natural spline fit of the cloglog-transformed predicted probabilities within a Cox model. The shaded area represents the 95\% confidence interval. (\textbf{B}) Binned calibration plot. Patients are grouped into 5 risk strata to compare average predicted versus observed event rates within each bin.} 
    \label{fig:calib}
\end{figure}

The main joint model demonstrates strong calibration, reflected by lower Integrated Calibration Index (ICI) values compared to the other two models, especially during the early 0–180 day period. The high-frequency joint model exhibits higher ICI values, suggesting calibration challenges likely driven by a smaller sample size, although it still maintains acceptable calibration overall. This underscores a fundamental tradeoff in our analysis: while incorporating more longitudinal data can enhance predictive accuracy, it may also reduce effective sample size and adversely affect calibration. Meanwhile, the benchmark model shows only moderate calibration, performing worse than the joint models, particularly on the test set. The main joint model's calibration in the 0-180 day period is visualized in Figure~\ref{fig:calib}. The smoothed calibration curve (panel A) shows a close agreement between the model's predicted probabilities and the observed event rates, as the solid blue line closely tracks the ideal dashed diagonal line. The 95\% confidence interval (shaded area) remains relatively narrow, suggesting stable calibration across the range of predictions. This finding is reinforced by the binned calibration plot (panel B), where all five risk strata (blue dots) lie very close to the line of perfect calibration, with their respective confidence intervals all overlapping the diagonal. Together, these plots confirm that the main joint model's risk estimates are reliable for clinical use.

\section{Discussion}\label{sec4}
This study demonstrates the value of integrating longitudinal NT-proBNP trajectories into a joint modeling framework to dynamically predict hospital readmission or death in patients with chronic heart failure. By leveraging real-world data and a Bayesian estimation approach, we show that modeling the full temporal evolution of NT-proBNP substantially enhances predictive performance compared to traditional static methods based solely on the most recent biomarker value. The joint model not only outperformed the benchmark model in accuracy, particularly at later time horizons, but also exhibited excellent calibration, supporting its reliability for individualized risk assessment.

A key strength of our work is the demonstration that dynamic prediction provides clinically relevant insights that static models cannot recover. For many patients, NT-proBNP trends during the early post-discharge period reflected meaningful changes in risk that were invisible when using a single measurement. Patients with decreasing trajectories showed correspondingly lower predicted risk, whereas increasing trajectories signaled elevated risk. This ability to update outcome predictions based on dynamic changes of the biomarker is one of the major advantages of joint modeling and highlights its potential utility in routine heart failure management. Our findings supplement and clarify prior work in the field: while prior research~\cite{zhang2018} suggested that the most recent NT-proBNP value might be sufficient for prediction, our results indicate that incorporating full trajectories leads to more accurate and adaptive risk estimates, particularly when additional follow-up information becomes available.

The clinical implications of these findings are substantial. Our model serves as a proof of concept for an advanced clinical decision support tool. Such a tool could provide clinicians with risk predictions that are updated in near real-time as new biomarker data becomes available. For example, a physician could see that a patient's risk of an adverse event in the next 180 days has increased based on a recent rise in their NT-proBNP trajectory, potentially prompting a change in management, such as a medication adjustment or a follow-up appointment. This moves beyond conventional and static risk scores and towards a more personalized, adaptive patient management strategy.

Despite the promising results, this study has limitations, primarily related to the nature of the real-world data used. Our reliance on a complete-case analysis for baseline characteristics and the requirement for a minimum number of biomarker measurements resulted in the exclusion of a significant portion of the initial patient cohort. 
%The final study population of 1,804 patients is only about 0.5\% fraction of the original 400,003. 
This creates a substantial risk of selection bias. It is plausible that patients with complete records and frequent follow-ups are treated at better-resourced healthcare centers, and may not be representative of the broader heart failure population. Consequently, the generalizability of our findings may be limited.
%While stringent filtering resulted in a substantially reduced cohort (1,804 patients), these criteria were necessary to guarantee high data quality and to enable robust estimation of longitudinal biomarker trajectories. 
Importantly, the objective of this study was to demonstrate the value of dynamic longitudinal modeling, and our results should be interpreted as a methodological proof of concept, rather than generalizable estimates for the broader TriNetX heart failure population. External validation with independent datasets will be necessary to assess the generalizability of these results.

While our model demonstrated the power of longitudinal data, many patients had the minimum required number of measurements (three), which constrains the ability to robustly model complex individual trajectories. We also treated baseline covariates as time-invariant, which is a simplification, as many of these factors can change over the course of the disease.
Furthermore, the model includes NT-proBNP measurements taken up to 30 days prior to the first discharge date due to HF. However, NT-proBNP levels can fluctuate significantly during hospitalization due to acute treatment. Our current analysis does not explicitly account for the in-hospital context, which may introduce noise into the initial trajectory estimates. Several avenues exist to address these limitations and strengthen the proposed framework in future work. Incorporating advanced missing-data strategies, such as multiple imputation, would enable inclusion of a larger and more representative patient cohort. Expanding the longitudinal component to jointly model additional biomarkers (e.g., creatinine, troponin) may further improve predictive accuracy. Additionally, accounting for competing risks could help differentiate between heart failure rehospitalization and all-cause mortality.
%Finally, refinements to the handling of pre-discharge biomarker variability could reduce noise in baseline trajectory estimation and enhance early predictions.
Such extensions would further refine individualized risk assessments and better support personalized treatment strategies. 

Although the real-world nature of the dataset inevitably imposed constraints, these challenges also highlight the robustness of the proposed framework. Even with heterogeneous measurement schedules, variation in clinical practice, and limited biomarker frequency for many patients, the model consistently delivered accurate and well-calibrated predictions. This reinforces the idea that joint modeling can be deployed effectively in practical healthcare environments and not merely in tightly controlled clinical trials. Furthermore, our use of a fully Bayesian approach supports uncertainty quantification at the patient level, enabling risk estimates that clinicians can interpret with an explicit understanding of their precision.

\section{Conclusion}\label{sec5}

In summary, our findings highlight the considerable promise of joint modeling as a tool for dynamic, patient-specific risk prediction in chronic heart failure. By leveraging the full NT-proBNP trajectory, the proposed framework provides more accurate, responsive, and clinically informative prognostication than traditional static approaches. As heart failure management continues to move toward personalized and proactive care, dynamic prediction models such as ours have the potential to serve as valuable components of future decision-support systems.

\backmatter

\section*{Supplementary Information}
\subsection*{Abbreviations}
\begin{tabular}{ll}
AFib & Atrial Fibrillation\\
BMI & Body Mass Index\\
CKD & Chronic Kidney Disease\\
COPD & Chronic Obstructive Pulmonary Disease\\
HF & Heart Failure\\
IBS & Integrated Brier Score\\
ICI & Integrated Calibration Index\\
JM & Joint Model / Joint Modeling\\
LVEF & Left Ventricular Ejection Fraction\\
MCMC & Markov Chain Monte Carlo\\
NT-proBNP & N-terminal pro-B-type natriuretic peptide\\
NYHA & New York Heart Association\\
SBP & Systolic Blood Pressure\\ 
\end{tabular}

\subsection*{Acknowledgements}
We thank Tobias Becher and Stanislas Hubeaux for their valuable feedback and constructive discussions on the manuscript.

\subsection*{Availability of data and materials}
Access to data from the TriNetX research network can be requested directly from TriNetX (\href{https://live.trinetx.com}{https://live.trinetx.com}). Approval is subject to a data sharing agreement and associated costs.

%\subsection*{Consent for publication}
%Not applicable.


\bibliography{sn-bibliography}% common bib file

\end{document}